\documentclass{aa}
\usepackage{psfig,graphicx}
\usepackage{txfonts}
\begin{document}
\title{Structure detection in the D1 CFHTLS deep field using accurate 
photometric redshifts: a benchmark  
}

\author{A. Mazure\inst{1} \and 
C. Adami\inst{1} \and
M. Pierre\inst{2} \and
O. Le F\`evre\inst{1} \and
S. Arnouts\inst{1} \and
P. A. Duc\inst{2} \and
O. Ilbert\inst{3} \and
V. LeBrun\inst{1} \and 
B. Meneux\inst{1,4,5} \and 
F. Pacaud\inst{2} \and 
J. Surdej\inst{3} \and 
I. Valtchanov\inst{6}
}

\offprints{C. Adami}

\institute{LAM, Traverse du siphon 13012 Marseille, France \and
DAPNIA/SAp, CEA Saclay, 91191 Gif sur Yvette, France \and 
Institut d'Astrophysique et de G\'eophysique, Universit\'e de Li\`ege, 
All\'ee du 6 Ao\^ut, 17, B5C, B-4000 Sart Tilman, Belgique \and
INAF - IASF, Via Bassini 15, I-20133 Milano, Italy \and 
INAF, Osservatorio Astronomico di Brera, Via Bianchi 46,
I-23807 Merate (LC), Italy \and 
Herschel Science Centre, ESA, European Space Astronomy Centre (ESAC), 
Villafranca del Castillo, PO Box 50727, Madrid 28080, Spain }

\date{Received 2006; accepted ????}

 
\abstract
{}
{We investigate structures in the D1 CFHTLS deep field in order to test the 
method that will be applied to generate homogeneous samples of clusters and 
groups of galaxies  in order to constrain cosmology and  detailed physics of 
groups and clusters.}
{Adaptive kernel technique is applied on galaxy catalogues. This technique
needs none of the usual a-priori assumptions (luminosity function, 
density profile, colour of galaxies) made with other methods. Its main 
drawback (decrease of efficiency with increasing background) is overcame by 
the use of narrow slices in photometric redshift space. There are two main
concerns   
in  structure detection. One is  false detection and  the second, the
evaluation of the selection function in particular if one wants  "complete"
samples.  We deal here with the first concern   using "random"
distributions. For the second,  comparison with detailed simulations is
foreseen but we use  here a pragmatic approach with comparing our results to
GalICS simulations to check that our detection number is not totally at odds
compared to cosmological simulations. We use XMM-LSS survey and secured 
VVDS redshifts up to z$\sim$1 to  check  individual detections.}
{We show that our detection method is basically capable to recover (in the
regions in common) 100$\%$ of the C1 XMM-LSS X-ray detections in the correct 
redshift range plus several other candidates.  Moreover when spectroscopic
data are available, we 
confirm our detections, even those without X-ray data.}
{}


\maketitle

\section{Introduction}

Considering groups and clusters of galaxies as "cosmological probes", two
questions are still presently asked: when these structures were formed and
could they constrain model-universes if we count them consistently with 
redshift and mass? Obviously, both questions point out for the need of 
"complete" samples of clusters and groups to answer them. Moreover, once 
such samples are available, detailed physics of galaxy groups and clusters 
can be studied.  

With the CFHTLS survey (Canada-France Hawaii Telescope Legacy Survey, see
http://www.cfht.hawaii.edu/Science/CFHLS) this could be done up to z$\geq$1 
extending e.g. recent GEMS study (Forbes et al. 2006) on nearby groups. 

Physically, groups and clusters are deep gravitational potential wells 
containing together dark matter, hot gas and galaxies. Each of these 
components could then in principle be used to detect their parent host.

a) Dark matter (DM hereafter): this phase is usually sampled via lensing 
studies (e.g. Gavazzi et al 2006 on the same CFHTLS fields and references
herein), since DM traces directly the mass. However,  
lensing detection is 
efficient up to not too high z systems (Hamana et al 2003), due to projection 
effects and lack of sensitivity to small groups. Moreover due to the complex 
mass distribution of DM haloes and the spoiling influence of intrinsic 
alignments of galaxies, lensing-selected clusters could  not be really 
mass-selected (e.g. Tang $\&$ Fan 2005).

b) Hot gas: this can be achieved in two ways. First, X-rays (e.g. Pierre et
al. 2006 and references therein) in principle
probe gravitational wells without projection effects, but are contaminated
by line of sight stars or active galaxies. Limited exposure
times also become problematic for faint remote clusters. 
 Second, the Sunyaez-Zeldovich (SZ hereafter) 
technique works at any redshift but in practice there are limitations for
distant objects due to the lack of spatial resolution. 

c) Galaxies: First, for this component, spectroscopic redshift surveys 
are in principle the optimal tool since they probe directly the dynamics 
of the systems. Most of the time, however, the spatial sampling is 
partial and not homogeneous. Moreover, high redshift clusters are not 
well probed due to slits/fibers overlap problems. Second, pure photometric 
catalogs can be used, searching for galaxy overdensities over the sky. This 
method provides in general an homogeneous spatial coverage, but the 
contrast of structures decreases rapidely with redshift with respect to 
the total background. Improvements are possible for example by  using  
matching filter technics (Postman et al 1996 for the seminal work, Olsen et al
2007 on the CFHTLS fields) and/or selecting  
galaxies on a colour-basis (using red sequence in the colour magnitude 
relation e.g. Gladders and Yee 2000, 2005)  but this method  is possibly introducing
a bias in the sense that it searches only for clusters exhibiting such a
  relation. Moreover, even if the colour magnitude relation already seems to
  be in place up to z$\sim$1.5 (e.g  Cucciati et al. 2006 for a complete
  discussion), the 
colour-magnitude distribution dependence upon environment also varies with
redshift. So, one would have ideally to adapt the cluster search using the
colour magnitude  
relation not only to the redshift but also to the local density.
This kind of improvement was also generalized in Miller et al. (2005),
  identifying clusters of galaxies as overdensities in a seven-dimensional
  position and color space to minimize the projection effects.

We chose a similar way to solve the problem of lack of 
contrast in using photometric redshifts to define redshift slices. But, up 
to now, the accuracy of such redshifts was moderate and significant numbers 
of totally wrong redshift estimates were still present.

However,
impressive improvements have been recently performed due to training of 
spectral templates and calibration with very good and large 
spectroscopic samples over large range of redshifts and down to similar 
magnitude depth compared to the photometric catalogs. We therefore used 
such improved photometric redshifts (Ilbert et al. 2006) in order to define 
redshift slices which allows to increase the contrast when computing 
galaxy density maps. This results in an homogeneous sample (besides masked 
photometric areas) only limited by the photometric catalog depth.

We took advantage of both excellent multi wavelength photometry from 
CFHTLS and very large samples of spectra (VVDS survey: VIMOS VLT Deep 
Survey, Le F\`evre et al. 2005) to define these very good photometric 
redshifts in the CFHTLS D1 field and, here, we exploit them to search 
for structures in that field. We postpone to another paper, in which all
the CFHTLS fields will be analysed, the use of counts of structures to 
constrain cosmological parameters. Indeed, only a few systems are expected 
in less than 1 square degree. Rather, we stress here on how the method 
appears efficient. In particular, a close comparison to the X-ray 
detections done in the frame of the XMM-LSS survey (e.g. Pierre et al. 
2006) shows a very good agreement when taking into account biases in both 
methods and we will also detail this comparison using secured VVDS redshifts.

Section 2 is about data and methods. Section 3 describes the structure 
detection and results reliability. Finally, Section 4 is the conclusion.

All along that paper we use the following cosmological parameters: 
H$_0$ = 67 km.s$^{-1}$.Mpc$^{-1}$,  $\Omega _\Lambda =0.67$ and 
$\Omega _m=0.33$ in order to be coherent with the GalICS simulations we used.

\section{Data and method}

\begin{figure*}
\centering
\caption[]{Masked regions following the CFHTLS recipe on the D1 field with the
  XMM detections  
superimposed. Red filled squares are the C1 XMM-LSS clusters and blue ones 
are the C2 and C3. We also give the name of the XMM-LSS cluster (see 
Pierre et al. 2006) as well as its redshift and bolometric X-ray 
luminosity in 10$^{44}$ erg/s within R$_{500}$. $\alpha$  and $\delta$ are
given in decimal degrees.} 
\label{fig:Fig1}
\end{figure*}

\subsection{CFHTLS photometric data and photometric redshifts}

The catalogues  (publicly available and fully described at 
http://terapix.iap.fr)  used for the detection and the  
characterization of the structures have been obtained within the frame of the 
CFHT Legacy Survey (i.e   MEGACAM u*, g', r', i', z' data) for the so-called
D1 deep field. We will use the i' band to detect our structures (see below). 

For the photometric redshift calculations, photometry (BVRI)  
obtained in the frame of the VVDS survey (Le F\`evre et al 2005) as well as 
Spitzer data are also used (see Ilbert et al. 2006 for details). 
 In a few words, photometric redshifts are obtained by adjusting Spectral
Energy Distribution of galaxy templates which are iteratively modified 
in terms of flux zero points and continua shapes using a set of high 
quality spectroscopic redshifts issued from the VVDS (see Ilbert et 
al 2006). A very good accuracy is obtained between z = 0.2 and z=1.2 and for
i' = 24.5 (the present limit of our sample, see below) as described in Ilbert
et al. (2006): in a photometric/spectroscopic redshift plot, the standard
deviation is 0.04 in redshift.

\subsection{The VVDS spectroscopic data  and XMM data}

 On one hand, once the structures are identified with their galaxy content
 (see following  
sections), we look for spectroscopic data in the VVDS catalogue both to 
confirm or not our detections and to give if possible an estimate of the 
velocity dispersion. However, the VVDS does not cover the entire D1 field and
is characterized by an inhomogeneous sampling rate and avoid peculiar masked regions
(distinct from the CFHTLS D1 masked regions however). So, several systems have
only a few or no spectroscopic measurements at all.  

On the other hand, the XMM-LSS (Pierre et al. 2006) provides for the CFHTLS D1
covered area, a catalog of candidate structures classified in several classes
(C1, C2, C3) with confirmed spectroscopic redshifts (independently from VVDS redshifts).  

Class 1 (C1) is defined as sources with no contamination of misclassified
point sources as extended ones. Class 2 (C2) corresponds to a contamination
of 50$\%$ and class 3 (C3) highly contaminated ones (see Pierre et al 2006 for
more details). We use also the identifications done by  
Willis et al. (2005) and Andreon et al. (2005) of specific systems with the
same X-ray data. 

First of all, Fig.~\ref{fig:Fig1} shows the  CFHTLS D1 masked regions due to
bright stars and CCD defects of the optical data. Masked regions are
represented by the spurious objects that were detected inside. Rings made by
agglomerated points with empty centers are for example due to stars that
shield part of the sky (i.e. empty centers) and contamine with their diffuse
light the immediate vicinity (e.g. rings of points).

It allows to check wether either an XMM 
source is found in a masked region and then if any optical detection 
could be spoiled by this masking (e.g. XLSSC 029 on the upper left very border 
of the D1 Megacam field). 
Second, we give in Fig.~\ref{fig:Fig3} the various XMM-LSS fields along 
with their corresponding associated exposure times (which can vary by a factor
of 2) along with the C1, 2 and 3 cluster detections. The part of the field at
both small right ascension and 
declination with no XMM observations at all will be called the ``Absence 
zone'' in the following.

\subsection{Galaxy density maps}

The method is based on the simple detection of contrasts in numerical density
maps of galaxies computed using i' band data.  But, in order to eliminate as much as possible fore  
and background contaminations, these  density maps are built in  redshift
(distance) slices.  
The technic used to compute the  galaxy density 
maps is the well known adaptive kernel method (see e.g Dressler and Shectman
1988,  Beers et al 1991 for  seminal works and also Biviano et al. 1996 for a
detailed application and discussion of significance).  
It has the advantage compared to wavelets (e.g. Escalera et al. 1992 or 
Slezak et al. 1994) of not needing reconstruction of the structures using the 
whole range of scales and is less affected by edge effects. Edge effects 
as well as mask effects (masks due for example to the presence of 
bright stars in the field) are not taken into account neither by mirroring the 
data nor by  adding randomly 
distributed points. 
We compared our detections with the map 
of masks and eyeballed if there was any unfortunate coincidence. Systems where
the center of the contours was in such a masked region were flagged by an (M)
in the lists of candidates (if not already associated with an X-ray structure).

In this 
testing approach, we prefer to deal with crude artefacts rather to smooth 
them in order to estimate their effects when compared to totally different 
detection methods. In an application aimed at counting groups and clusters 
to constrain cosmology, it turns out that the best way would be to exclude
totally and a posteriori masked regions as done e.g when using lensing 
techniques. For the edge effects, comparison of our detections with real 
structures detected in X-ray will indicate in the following that the effect 
does not seem to be so important. Again, when counting structures, this should
be taken  
into account by removing adequately border zones. Only foreseen comparison 
with simulations including fake structures where completeness in terms of 
richness or mass is well controlled, will allow quantitative estimates of
these effects.   

We thus define overlapping slices with width of 0.1 in photometric
redshift all along the line of sight. An overlap of 0.05 is chosen as real 
structures are often expected to exhibit in adjacent slices.

The D1 field is 0.8 deg$^2$ (after masked area rejection) and we define on it 
a grid of 200$\times$200 pixels. The pixel size ($\sim$0.3 arcmin) 
corresponds to $\sim$80/120/160 kpc at z=0.25/0.5/1 and is let fixed with 
redshift.
  
In order to establish statistical significance, we use bootstrap technique
both on real data and  simulations. For each new 
realization  of a  given galaxy distribution obtained by the bootstrap
technique we build the corresponding  density map.  In every new realization
of the galaxy distribution, clustered  
points at "small" scales stay clustered but are hovewer spread over. Points  
unclustered or clustered on larger scales are also spread over.  Then, taking
the  
mean of several density maps   has   the effect to erase fluctuations and
flatten the mean background,  
letting clustering present. This flattening added to the removal of distant 
(or nearby) clustering due to the use of narrow slices allows the use of random 
distributions to evaluate false detections at least when using high value 
thresholds. 
In practice  a "mean bootstraped" map of the galaxy distribution within a
given zphot slice is obtained using 1000 bootstrap resamplings (see e.g.  
Biviano et al. 1996 for a complete description).

\section{Application to real data and Reliability}
  
\subsection{Application to real data}
   
   The present  analysis is performed using the i' band CFHLS catalogue, down
   to the  
magnitude limit i' = 24.5 which encompasses the typical i'* of the luminosity
function at z=1 (e.g. Adami et al 2005)  and  which ensures  a good compromise
between going as red as possible and using the best  quality   photometric
data.     

We apply the technique to overlapping slices with central redshifts between
0.2  and 1.2 (where the quality of photometric redshifts is optimal).  
We add the slice 0.10-0.20 to check if the nearby C1/0.041 is
detected by our method. We also give in appendix the tentative detections up
to z=1.5, but these last results are still uncertain due to the decrease of the
photometric redshift accuracy after z=1.2. This is also why the last redshift
slice in appendix was taken larger (width of 0.15 in redshift) than the other ones. 

Once for a given slice, the mean  (from  1000 realizations 
obtained by bootstraping the actual data) "image" (i.e the mean galaxy density map) is
obtained and a detection of the  
density peaks is performed on it.  It uses the usual image analysis done with 
Sextractor (Bertin $\&$ Arnouts 1996) where the internal parameters are
adapted to the pixel size with at least 2 pixels above the choosen threshold
(3 or 4). As in usual image analyses, structures are detected 
with respect to their background (estimated globally in Sextractor) which
defines a threshold (peak density over background density).  
Positions and best ellipse fitting (orientation  and axes ratio) are derived 
adopting 1 Mpc as the size of the semi-major axis of the considered structures.    

A catalogue of structures for each redshift slice is then generated. We
also select individual galaxies potentially belonging to each structure as 
all galaxies included in the considered redshift slice and in the 1 Mpc semi-major axis
ellipse. For a given structure, these lists still include, however, interlopper galaxies 
that have positions inside the structure ellipse but are foreground or
background galaxies included in the redshift slice. This is due to 
the photometric redshift uncertainty (see Ilbert et al. 2006). 
    
We avoid the very low redshift slices because in that case, the
number of expected structures is small due to the small  
solid angle and the reliability of photometric redshifts is degraded 
(see Ilbert et al. 2006). We also provide detections only up to z=1.2
(tentative detections up to z=1.5 are only given in appendix), but we 
clearly wait for complete comparison with the next generation of 
simulations to validate and study detections of the most distant candidate clusters in 
terms of mass.  As we use galaxy as tracers of structures,  it is important to
deal with large scale structure simulations  which  include a  well controled  implementation
of galaxies and not only of DM. It is however encouraging that the number of
detections  agree  well with generic GalICS predictions (see below). 
  
Every significant  (i.e in terms of threshold, see below) structure  is
labelled with an identification in every slice as well as a general identification. 
Some structures could show up at almost the same positions in several
slices. In order to identify these multiple detections, we chose to give a
single identification when two detections in adjacent redshift slices had
overlapping ellipses on the sky. We show in Fig.~\ref{fig:shiftref} the coordinates differences
of all multiple detections in successive redshift slices. This figure shows
that two successive detections always are closer than 2Mpc (by definition) and
that more than 75$\%$ are closer than 1Mpc. The values are of the order (or
smaller) than 2 times the usual virial radius and this ensures that most of
the time, we are not merging unrelated structures. The other 
detections of a given structure are then labelled by the same number but
flagged by parentheses in Tabs.~\ref{tab:candi} to ~\ref{tab:candi2} which 
give a summary of the structures we found. 

We note, however, that when a structure (e.g. large scale structure
line-of-sight filaments) is percolating through a large number of redshift
slices, the position of the lowest redshift detection can be quite different
with the one in the highest redshift slice.
We also note that C1-029 from Pierre et al. (2006) is just a the edge of the D1
field of view and is also located in a masked region. It is perhaps identified
with the cluster 35 (general id of the tables), but this remains very uncertain.

Finally, as part of the D1 field is covered by VVDS data, we use the 
spectroscopic information. In every 1 Mpc ellipse, when available, we look at 
spectroscopic data in the 0.1 width redshift range to confirm if the local z
distribution exhibits any compactness in the velocity space compatible with the presence of  
possible real structures. Namely, we looked at galaxies along the line of
sight in the considered slice separated by gaps of less than
0.0026 (in order to use the same gap as in Adami et al. 2005 on similar
spectroscopic data). These gaps were adapted to redshift using the (1 + z) dependence.

\subsection{False detection rate evaluation}
  
The first and main concern with any detection method is the question of false 
detections. Here we deal with that in the following way:

We generate 100 independent slices with 5000, 8000 and 11000 points
randomly distributed (these numbers are representative of numbers of galaxies
in real redshift slices as shown in tabs.~\ref{tab:candi} to
~\ref{tab:candi2}) and we analyze these slices in the same way as real ones. 
We compute the number of detections depending on the detection thresholds 
used.  
 
As we proceed in narrow slices, the effect of distant clustering which  
diminishes the contrast of real structures is also strongly diminished 
as well as with the bootstrapping. So, mean random fields are rather good 
representations of the actual slice background.   

Table~\ref{tab:area2} shows the numbers of detections in a given slice (whatever
the redshift) with respect to the threshold defined in terms of 3 and
4$\sigma$ of the (local) background. This table, then, gives an estimate of the 
level of wrong structure detections for a given slice. 
This level remains modest, since there is  at most 2 in a given slice  at 
the   3$\sigma$ level  and 1 at the 4$\sigma$ level.

\begin{table}
\caption[]{Number of peaks detected per square degree in 100 random fields 
(with 5000, 8000 and 11000 points each) with respect to a given threshold.}
\begin{tabular}{cccc}
\hline
Threshold  & 5000 points & 8000 points & 11000 points  \\
\hline
$\geq$3-$\sigma$  &  0.7$\pm$0.8 & 0.6$\pm$0.6   & 1.6$\pm$0.9  \\
$\geq$4-$\sigma$  &  0.1$\pm$0.2 & 0.2$\pm$0.4   & 0.6$\pm$0.6  \\
\hline
\end{tabular}
\label{tab:area2}
\end{table}

\subsection{Global detection rate assessment with Galics simulations and other 
optical detection methods}

As a first step we generate, using the 50 available GalICS simulations (with H$_0$ = 
67 km.s$^{-1}$.Mpc$^{-1}$,  $\Omega _\Lambda =0.67$ and $\Omega _m=0.33$), 
slices with the same widths as the ones with real data (e.g. Meneux et al. 
or Blaizot et al. 2006 for a discussion on the ability of these simulations 
to represent the real universe). These simulations are representative of 
the general clustering in the Universe all along with the same depth as our 
sample. For each GalICS slice we produce mean bootstraped maps in the same 
conditions as for real data. In the present stage (see below), we will just 
check with these simulations that the used thresholds lead to a number 
of detections (see Tab.~\ref{tab:area}) not totally at odds with the ones provided  
by the real fields. 
Of course, as well in real data as in simulated ones, false detections (due
e.g to projection effects) and cosmic variance affect the number of
structures. Consequently, numbers are expected to agree only in the mean and
we recall that a complete evaluation in terms of 
richness, mass and other characteristics is devoted to another paper. 

\begin{table}
\caption[]{Number of peaks detected per square degree in 50 GalICS fields 
(averaged over the 50 available fields) and in real D1 fields (rescaled to 1 deg$^2$) 
with respect to redshift and to a given threshold. 
We only give the non-overlapping redshift slices.}
\begin{tabular}{ccccc}
\hline
 z  & $\geq$3$\sigma$ Gal& $\geq$4$\sigma$ Gal & $\geq$3$\sigma$ D1 & 
$\geq$4$\sigma$ D1 \\
\hline
 0.2-0.3  & 7.0$\pm$0.8 &  6.2$\pm$0.8 & 7.5 &  5.0\\ 
 0.3-0.4  & 8.0$\pm$0.8 &  6.8$\pm$0.8 & 5.0 &  3.75\\ 
 0.4-0.5  & 8.1$\pm$0.9 &  6.8$\pm$1.0 & 5.0 &  1.25\\ 
 0.5-0.6  & 8.3$\pm$0.9 &  6.6$\pm$1.0 & 7.5 &  3.75\\ 
 0.6-0.7  & 8.0$\pm$0.9 &  6.0$\pm$1.1 & 2.5 &  1.25\\
 0.7-0.8  & 7.6$\pm$1.1 &  5.4$\pm$0.7 & 5.0 &  1.25\\
 0.8-0.9  & 6.9$\pm$0.7 &  4.7$\pm$0.8 & 6.25 &  2.5 \\ 
 0.9-1.0  & 6.3$\pm$0.6 &  4.3$\pm$0.6 & 6.25 &  2.5\\ 
 1.0-1.1  & 5.6$\pm$0.6 &  3.6$\pm$1.0 & 3.75 &  2.5\\ 
 1.1-1.2  & 4.9$\pm$0.6 &  3.0$\pm$0.5 & 5.0 &  2.5\\ 
 1.2-1.3  & 5.4$\pm$0.5 &  3.2$\pm$0.6 & 6.25 &  5.0\\ 
\hline
\end{tabular}
\label{tab:area}
\end{table}

We, however, compare from Tab.~\ref{tab:area} and fig.~\ref{fig:galD1} 
the number of detections in
our sample and in the GalICS simulations. This will give a global estimate 
of how well our detections are in agreement with the cosmological model 
used in the chosen GalICS simulations. Fig.~\ref{fig:galD1} shows that 
our number of detections is in good agreement with the ($\Omega _\Lambda$ 
=0.67 and $\Omega _m$=0.33) model given the error bar sizes. 

Of course these detections could be in both cases as well real or false 
detections. However, as seen above using the random fields representative 
of each slice background, the high level thresholds used show that the 
false detection rate is expected to be rather low. It would be an unlikely 
coincidence that in every slice, the number of real and false detections 
conspire to give the right numbers found. 

We, however, note that we detect less structures at intermediate redshifts  
(0.55;0.75) than predicted by the model. But this corresponds also (see 
below) to a lack of X-ray detections revealing that it could be a real 
empty region. Finer 
comparisons are required in order to really constrain cosmology, but the
relatively good agreement between simulations and observations gives
confidence on the potentialities of the method.

As seen from tables~\ref{tab:candi} to ~\ref{tab:candi2}, we 
detect, up to z=1.1 about 40 independent structures within an area of 
0.8 deg$^2$ which is very close to the 52$\pm$8 per deg$^2$ found by Olsen 
et al. (2007) using a totally different method and a slightly deeper magnitude
limit (but given the considered magnitudes, this should only affect the faint 
population of our detected structures and not too strongly the structures
themselves). A close comparison between the two methods is devoted to another paper.

\begin{figure}
\centering
\caption[]{Regions observed by XMM with the corresponding C1, C2 and C3
detections as in Fig.~\ref{fig:Fig1} (see Pierre et al 2006). The red
circles have a shorter X-ray exposure time. Red filled squares are the 
C1 XMM-LSS clusters and blue ones are the C2 and C3.}
\label{fig:Fig3}
\end{figure}

\begin{figure}
\centering
\mbox{\psfig{figure=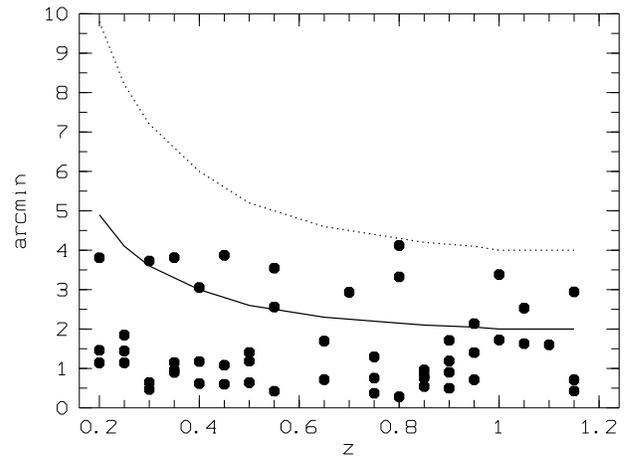,width=9cm,angle=270}}
\caption[]{Identification distance in arcmin between two detections (in
  successive redshift slices) that are assumed to be the same structure as a
  function of redshift. Solid and dashed lines are (in arcmin) the 1 and 2 Mpc 
  values as a function of redshift.}
\label{fig:shiftref}
\end{figure}

\begin{figure}
\centering
\mbox{\psfig{figure=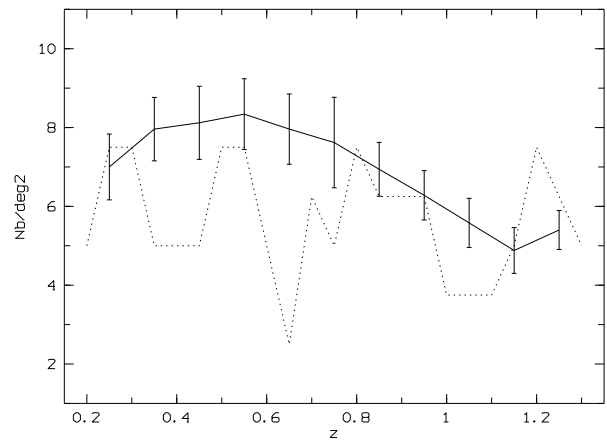,width=9cm,angle=270}}
\caption[]{ Number of 3$\sigma$ detections per deg$^2$ in the GalICS 
simulations: continuous line with error bars. Number of detections per 
deg$^2$ in the D1 field: dashed line. }
\label{fig:galD1}
\end{figure}

\begin{figure}
\centering
\caption[]{Slice 0.10-0.20 (i.e. 0.10$\leq$z$\leq$0.20). Colour contours are 
drawn from the mean galaxy 
density image. Bottom level is the mean value of this density map. Black 
small dots show the fake galaxies detected in the D1 masked areas. Large 
circles are the XMM-LSS fields. Peaks are detected at the 3$\sigma$ level 
(check tables of structures to see which ones are also detected at the 4$\sigma$ 
level) and can be distinguished by their number. C1 XMM clusters (from Pierre
et al. 2006) in the same redshift range (with an allowance of 0.01) are 
superimposed as red squares, C2 and C3 (from Pierre et al. 2006) as white
squares. Coordinates are given in decimal degrees (J2000).}
\label{fig:Fig010020}
\end{figure}

\begin{figure}
\centering
\caption[]{Same as ~\ref{fig:Fig010020} for the slice 0.15-0.25.}
\label{fig:Fig0.15-0.25}
\end{figure}

\subsection{Comparison with XMM-LSS and the VVDS}

As we already said, the second concern (once false detection 
problem is addressed) is the ability of any method to detect real 
structures at the right place and at the right distance or rather to
understand when and why they are not detected. In other words what is the 
so-called detection function associated? 

One way to estimate correctly such a selection function is the use of
cosmological simulated catalogs analysed with exactly the same protocol as 
for the real data. One can also use mock catalogs in which structures are 
put by-hand allowing the recovery power estimation. However, if it is 
pedagogical in the sense that it gives a flavour of how the algorithm 
works, it is not always possible to explore all the range of parameters and 
conditions in a practical way until detailed cosmological simulations adapted
to the CFHTLS characteristics will be available. 

A second way to estimate such a selection function is empirical: the
use of detections obtained in a totally different way (X-ray emission 
of groups and clusters detected by the XMM-LSS) accounting in the meantime 
for spectroscopic information issued from VVDS. This spectroscopic 
information is used in the following way. For every structure in a given 
slice, we look at the redshift distribution of the VVDS data  along the line
of sight, within  
a 1 Mpc ellipse, up to z =1.2 (central redshift of the highest redshift
considered slice) and we look for clustering in spectroscopic  
redshift within the redshift range of the slice. We put galaxies as
possible members of a system when the redshift of 2 galaxies do not differ
by more than 0.0026$\times$(1 + z) (see also Section 3.1). All the 
corresponding numbers of this analysis are given in table~\ref{tab:ZSPEC}. 
  
Here, we check if, with the thresholds defined above, we recover or not   
the various XMM sources. Of course, there are physical reasons for not 
detecting in X-ray an optical overdensity (e.g. if dealing with a non 
totally virialized system) as well as observational ones since XMM-LSS does 
not cover entirely D1 and has not a completely homogeneous exposure time. 
Conversely, if there is no optical structure when a diffuse X-ray 
source is present, it could reveal a failure of the present method or a 
mistake in its distance as well as a peculiar class of extended X-ray
structures (as fossil groups that are galaxy-very-poor massive systems,
e.g. Jones et al. 2003 or Ulmer et al. 2005).

\begin{figure}
\centering
\caption[]{Same as ~\ref{fig:Fig010020} for the slice 0.20-0.30.}
\label{fig:Fig0.20-0.30}
\end{figure}

\begin{figure}
\centering
\caption[]{Same as ~\ref{fig:Fig010020} for the slice 0.25-0.35.}
\label{fig:Fig0.25-0.35}
\end{figure}

\begin{figure}
\centering
\caption[]{Same as ~\ref{fig:Fig010020} for the slice 0.30-0.40.}
\label{fig:Fig0.30-0.40}
\end{figure}

Figures~\ref{fig:Fig010020} to~\ref{fig:Fig1.00-1.10} show a subsample of 
the slices up to z = 1.1 for thresholds of 3$\sigma$. We note that these
figures are plotted with the same color coding, with the lowest level being
the mean value of the density map. If different slices exhibit strongly
different background noise (or different maximal density values), then,
significant peaks can appear less proeminent in the figures compared to less
significant peaks in slices with different noise level. The individual 
properties of the structures are given in Tabs.~\ref{tab:candi} to 
~\ref{tab:candi2}. 

We overplotted XMM-LSS clusters in the range 0.1 - 1.2 with the
following rule: C1, C2 and C3 XMM-LSS clusters were plotted in our
graphs when included in the considered photometric redshift range (with 
an allowance of 0.01, only useful for C1-025). 

In the tables, when a given structure is detected in adjacent slices, to identify it 
with an X-ray source, we always select the one corresponding to the highest 
significance level. The X-ray position has to be included in the optical
detection ellipse. Finally, in the tables, we also take into 
account the XMM-LSS clusters provided by Andreon et al. (2005) and Willis et 
al. (2005) in particular analyses, separate from Pierre et al. (2006), but
using the same XMM data.

\begin{figure}
\centering
\caption[]{Same as ~\ref{fig:Fig010020} for the slice 0.35-0.45.}
\label{fig:Fig0.35-0.45}
\end{figure}

\begin{figure}
\centering
\caption[]{Same as ~\ref{fig:Fig010020} for the slice 0.40-0.50.}
\label{fig:Fig0.40-0.50}
\end{figure}

\begin{figure}
\centering
\caption[]{Same as ~\ref{fig:Fig010020} for the slice 0.45-0.55.}
\label{fig:Fig0.45-0.55}
\end{figure}

\begin{figure}
\centering
\caption[]{Same as ~\ref{fig:Fig010020} for the slice 0.50-0.60.}
\label{fig:Fig0.50-0.60}
\end{figure}

\begin{itemize}

\item First, it should be noted that for redshift higher than 0.1 
all XMM-LSS C1 clusters are detected at the 3$\sigma$ level with the
present method (except for C1-029 almost outside the optical field 
and in a masked region of this field). Most of these are also detected at the
4$\sigma$ level except C1-005 (detected at the 3.75$\sigma$ level) and C1-025
(at 3.3$\sigma$ level).

\item C2-038 is not detected in our analysis and is neither in a masked
 nor in an edge region. Looking in more detail at the maps shows 
that it is detected at only a 1.5$\sigma$ level in the [0.50;0.60] slice. 

\item We also detect 2 out of the 4 C3 clusters. Surprinsigly, the C3 
detected are at high z while the non-detected ones are at low redshift. 
However, C3-a appears close to a bright star and is probably spoiled 
in the CFHTLS data (masked region). For C3-d, it appears also to be detected at 
only 1.5$\sigma$ in the [0.30;0.40] slice.

\item We finally find coincidence in the correct 
redshift range (see Tabs.~\ref{tab:candi} to ~\ref{tab:candi2}) for structures 
4 and 34 with the X-ray sources confirmed by Andreon et al. (2005) and Willis 
et al. (2005).

\item It must be underlined that our analysis succeed  to recover afterwards
the cluster 0004 (number 15) of the [0.50;0.60] redshift slice, identified as 
a faint X-ray source but rejected as a possible extended source from the XMM-LSS
list in a first stage (so not included in the C1/C2/C3 classification). 
This X-ray source is very close to the bright 
XMM-LSS C1 cluster 041 but is still detected by our method (see 
Figs.~\ref{fig:Fig0.45-0.55} and ~\ref{fig:galD2}). This illustrates 
the ability of our method to efficiently disentangle nearly superposed 
clusters.  We also found detections apparently without any X-ray 
identification. One clear example is given by structure 12 
detected here at the 3$\sigma$ level without confirmed X-ray counterpart in
the redshift bin.  For this region, 
however, more than 30 VVDS redshifts are available confirming real clustering 
at z$\sim$0.31.

\end{itemize}
 
In summary, there are 1 C2 cluster and 2 C3 clusters that are not detected at the 
3-$\sigma$ level (but 2 are recovered at lower levels) with no obvious
explanation among the confirmed clusters from
Pierre et al. (2006). This corresponds to a level of $\sim$15$\%$ which turns
out to be the number of completely missed structures by our method compared to X-ray 
methods.
 
Conversely, as said above, several structures are detected in the visible at a
significant level with no counterparts in X-ray. Are they real or are they
false detections? Restricting to areas in common, with no biases (i.e 
excluding A,G,E,M,S regions: see Tabs.~\ref{tab:candi} to ~\ref{tab:candi2}) 
and limiting ourselves to z$\leq$1.05 (the limit in redshift of Pierre et
al. 2006) we find 11 such structures of which 5 have spectroscopic
information. Namely these are systems 2, 16, 19, 21 and 28 (see
Tab.~\ref{tab:ZSPEC}) and are therefore likely to be real. We also note that
structure number 3 (general id) is detected at z=0.26 by Pierre et al. (2006)
and at z=0.225 using only VVDS redshifts. This discrepancy is probably due to
the small number of VVDS redshifts (4) in the considered slice leading to a
wrong estimate.

We conclude that we are probably more 
efficient to detect very low mass and galaxy-dominated systems (as compared 
with gas or dark matter dominated systems) compared to X-ray methods. These 11
only optically detected structures is the number of probably missed structures 
by the X-ray method.

\begin{figure}
\centering
\caption[]{Same as ~\ref{fig:Fig010020} for the slice 0.55-0.65.}
\label{fig:Fig0.55-0.65}
\end{figure}

\begin{figure}
\centering
\caption[]{Same as ~\ref{fig:Fig010020} for the slice 0.60-0.70.}
\label{fig:Fig0.60-0.70}
\end{figure}

\begin{figure}
\centering
\caption[]{Same as ~\ref{fig:Fig010020} for the slice 0.65-0.75.}
\label{fig:Fig0.65-0.75}
\end{figure}

\begin{figure}
\centering
\caption[]{Same as ~\ref{fig:Fig010020} for the slice 0.70-0.80.}
\label{fig:Fig0.70-0.80}
\end{figure}

\begin{figure}
\centering
\caption[]{Same as ~\ref{fig:Fig010020} for the slice 0.75-0.85.}
\label{fig:Fig0.75-0.85}
\end{figure}

\begin{figure}
\centering
\caption[]{Same as ~\ref{fig:Fig010020} for the slice 0.80-0.90.}
\label{fig:Fig0.80-0.90}
\end{figure}

\begin{figure}
\centering
\caption[]{Same as ~\ref{fig:Fig010020} for the slice 0.85-0.95.}
\label{fig:Fig0.85-0.95}
\end{figure}

\begin{figure}
\centering
\caption[]{Same as ~\ref{fig:Fig010020} for the slice 0.90-1.00.}
\label{fig:Fig0.90-1.00}
\end{figure}

\begin{figure}
\centering
\caption[]{Same as ~\ref{fig:Fig010020} for the slice 0.95-1.05.}
\label{fig:Fig0.95-1.05}
\end{figure}

\begin{figure}
\centering
\caption[]{Same as ~\ref{fig:Fig010020} for the slice 1.00-1.10.}
\label{fig:Fig1.00-1.10}
\end{figure}

\subsection{Clusters, groups and filament properties}

\begin{figure}
\centering
\mbox{\psfig{figure=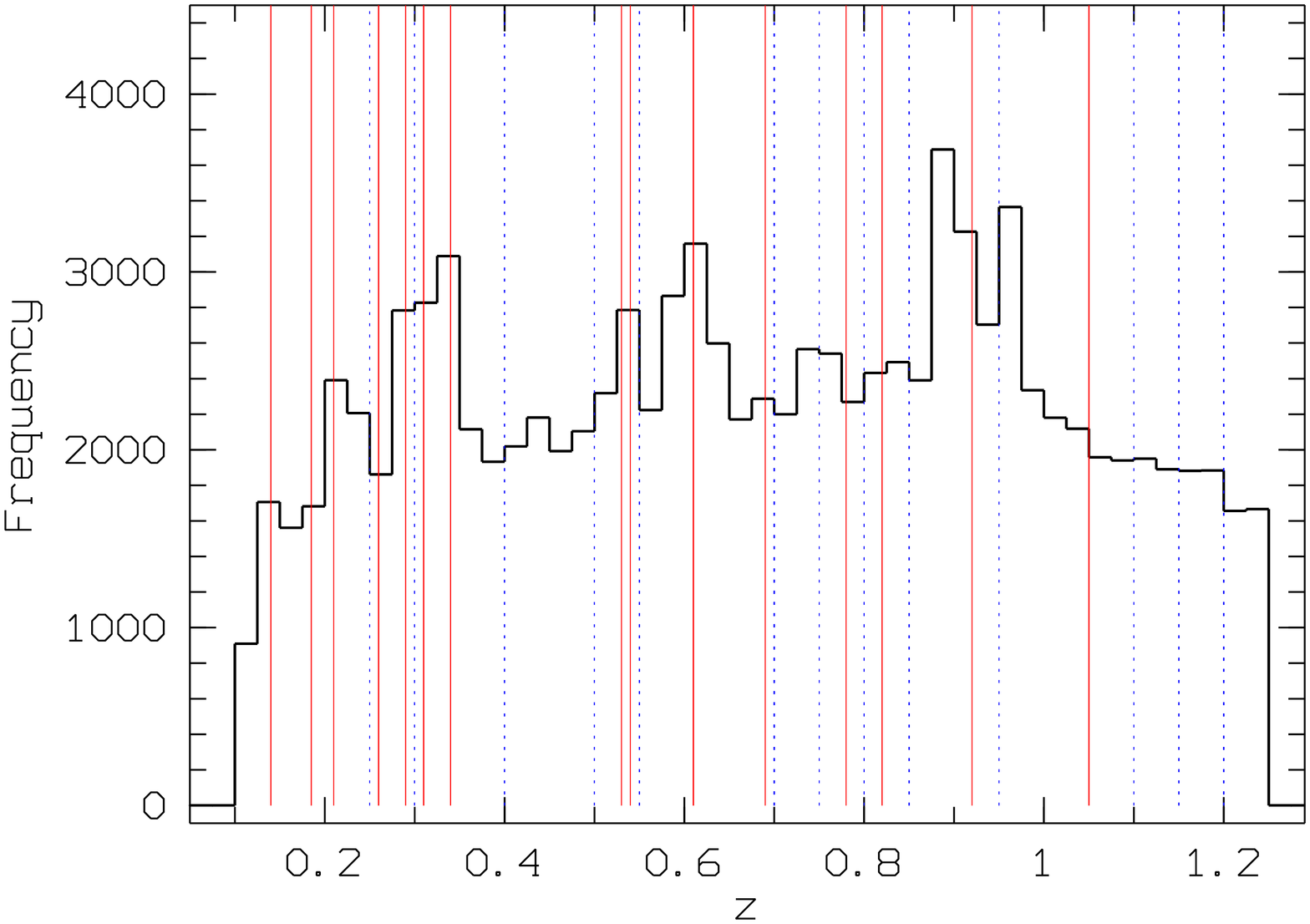,width=9cm,angle=270}}
\mbox{\psfig{figure=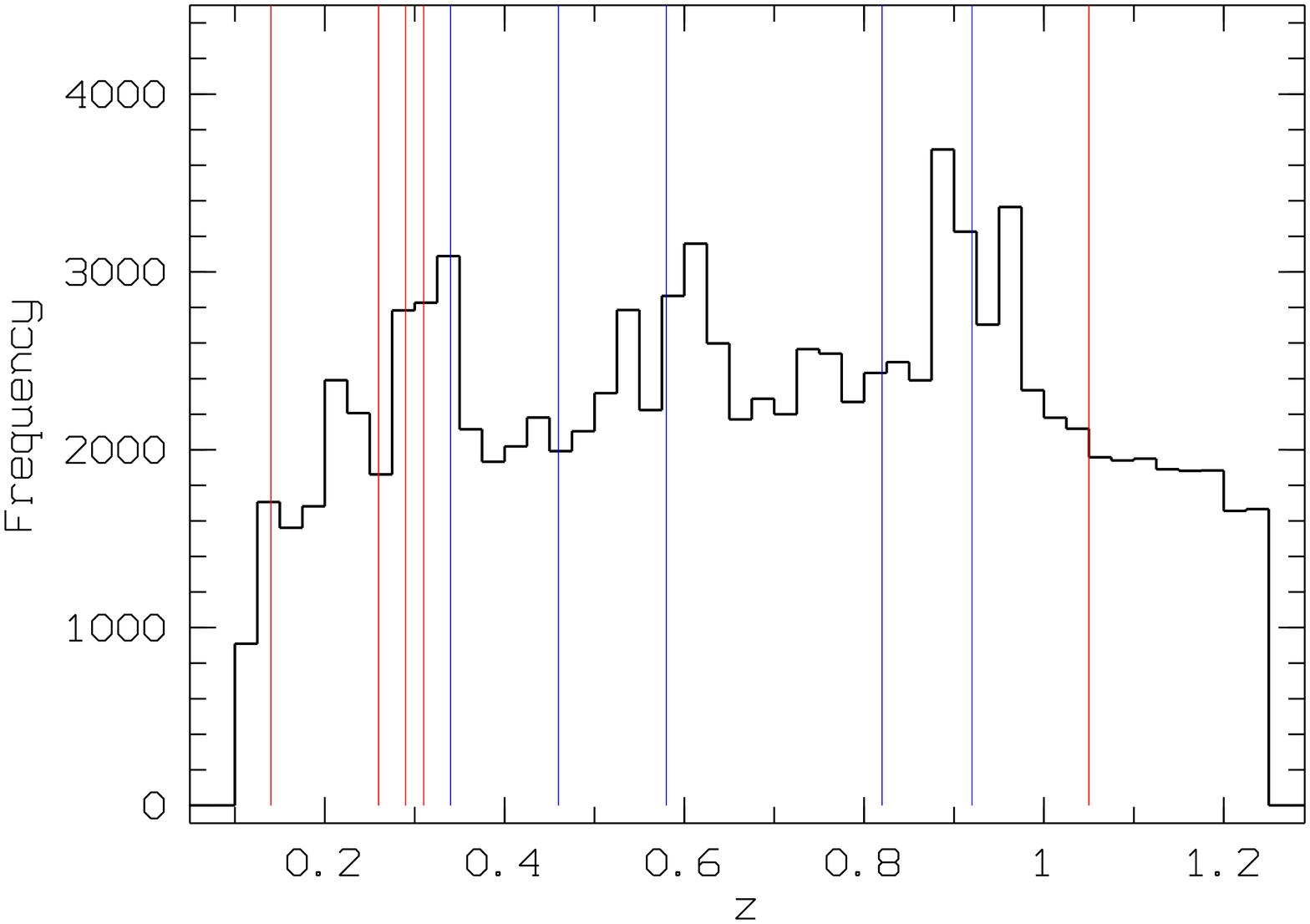,width=9cm,angle=270}}
\caption[]{Histogram of
  all photometric redshifts along the line of sight between z=0.1 and 1.25
  overplotted with detected structures. 

  Upper figure: structures detected with
  the photometric redshifts. Those with a precise redshift
  determination (first from XMM-LSS papers and second from VVDS spectroscopic
  data) are the red continuous lines. The blue dashed lines are clusters with
  only the photometric redshift determination (taken as the central redshift
  of the considered slice).
 
  Lower figure: structures detected in X-rays from Pierre et al. (2006). 
  C1 clusters are the red lines. C2 and C3 clusters are the blue 
  lines. }
\label{fig:historef}
\end{figure}

A future paper will be dedicated to the precise study of the properties of 
these structures but we show in Fig.~\ref{fig:historef} the histogram of
  all photometric redshifts along the line of sight between z=0.1 and 1.25
  overplotted with detected structures. We clearly see that we detect
  structures in almost all galaxy concentrations in the redshift space. We
  also detect several structures in low density regions.

\begin{figure*}
\centering
\mbox{\psfig{figure=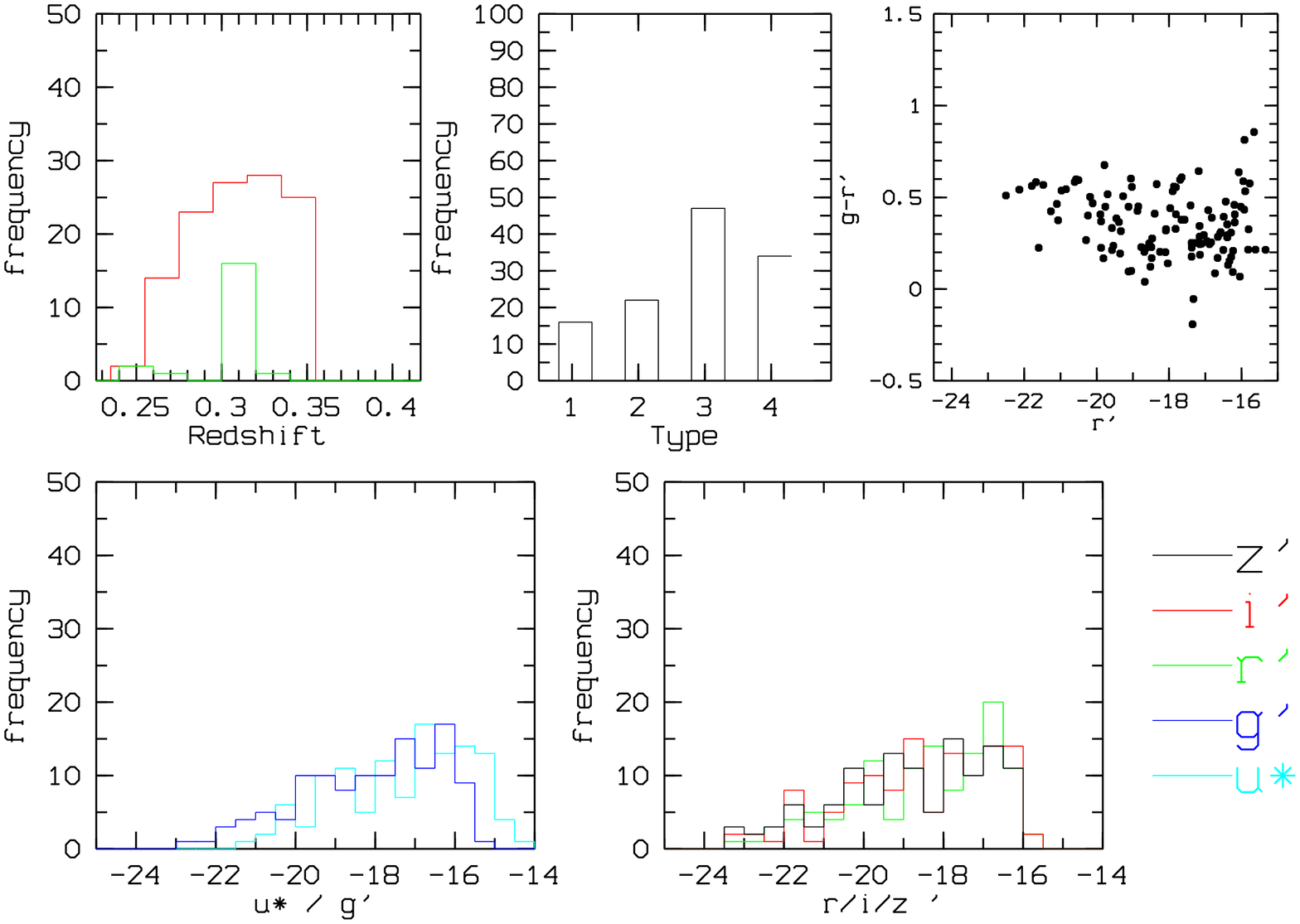,width=18cm,angle=270}}
\caption[]{Summary of the properties for the candidate 0004 in the [0.25;0.35] 
redshift slice. Lower left: u* and g' band absolute magnitude histograms, 
lower right: r', i', z' absolute magnitude histograms, upper left: 
redshift histograms (photometric redshifts: red, spectroscopic redshifts: 
green), upper center: spectrophotometric types (following Coleman et
al. 1980), upper right: colour magnitude relation.}
\label{fig:example}
\end{figure*}

We also present an example of what can be done. 
Fig.~\ref{fig:example} shows for example the luminosity functions in the 
CFHTLS u*, g', r', i' and z' bands of the candidate 0004 in the [0.25;0.35] 
redshift slice. Objects are selected in the slice (so we still have some
foreground and background contamination by galaxies in the considered slice
but not physically included in the structures) and within the 1 Mpc ellipse.
It also shows the redshift and spectro-morphological type 
histograms (following Coleman et al. 1980) as well as the red sequence 
in the Colour Magnitude Relation. 

\begin{figure}
\centering
\caption[]{Trichromic r/i/z CFHTLS image of candidate 0004 in the slice 
0.50-0.60. The three XMM-LSS spectroscopic redshifts (distinct from the 2 VVDS
redshifts) are shown. The 
large galaxy to the top of the image is related to the XMM-LSS C1 cluster 
041 at z=0.14.}
\label{fig:galD2}
\end{figure}

\begin{table*}
\caption[]{From a central redshift of 0.15 to 0.70.
Structures detected along with their redshift slice (and the total number of
 galaxies inside the slice), structure id, 
coordinates, minimum detection level, general identification, X-ray association
(redshift coming from Pierre et al. 2006, Willis et al. 2005 or Andreon et al. 2005)
or (if not identified in the slice or in another slice) reason 
potentially explaining the absence of X-ray 
detection (center of the structure located in A: the Absence zone, G: 
a Gap between two X-ray pointings, M: a region 
strongly affected by Masked CFHTLS areas, S: a single XMM field 
where the exposure time was Short, E: a region with an X-ray detection very 
close to 
the CFHTLS D1 field Edge), X-ray temperature when available and VVDS redshifts
available (Yes or No). Some structures
show up at almost the same positions in several slices: the secondary 
detections are then labelled by the same number but flagged by parentheses
(the unflagged number is the detection made with the highest signal to noise). 
W05 refers to Willis et al. (2005) and A05 to Andreon et al. (2005).}
\begin{tabular}{ccccccccc}
\hline
Slice &  id  & $\alpha$ & $\delta$ & Min. thres.
& gen. id. & X-ray id and z XMM & T-X  & VVDS  \\
\hline
0.10-0.20 (5868) & 0001 & 36.3789 & -4.2424  & 4 & 1  & C1-041/0.14 & 1.3 keV &Y\\
         & 0002 & 36.7981 & -4.1970  & 3 & 2  &             &     &Y    \\
\hline
0.15-0.25 (7848) & 0001 & 36.3746 & -4.6831 & 3  & 3 & C1-025/0.26 & 2.0 keV &Y\\
         & 0002 & 36.5767 & -4.0699 & 3  & (4) &  S           &      &N  \\
         & 0003 & 36.6240 & -4.2523 & 3  & 5 &  S           &    &Y   \\
         & 0004 & 36.8215 & -4.5465 & 3  & (6) & S &  &Y  \\
\hline
0.20-0.30 (9252) & 0001 & 36.8946  & -4.8679  & 4  & 7 & C1-022/0.29 & 1.7 keV&N\\
         & 0002 & 36.2451  & -4.8901  & 3  & 8 & A/M           &     &N   \\
         & 0003 & 36.1364  & -4.2134  & 4  & 9 & C1-044/0.26 & 1.3 keV& \\
         & 0004 & 36.6840  & -4.2315  & 3  & (5) &        &        &Y \\
        & 0005 & 36.8435  & -4.5570  & 4  & (6) &  & &Y\\
         & 0006 & 36.5928  & -4.0801  & 4  & (4) & S            &     &N  \\
\hline
0.25-0.35 (10570) & 0001 & 36.3166 & -4.7515 & 3  & 10 & MA& &Y \\
         & 0002 & 36.9117 & -4.8594 & 4  & (7) &  &  &N \\
         & 0003 & 36.6381 & -4.8759 & 4  & 11 & M           &       &N \\
         & 0004 & 36.8416 & -4.5810 & 4  & 6 & C1-013/0.31 & 1.0 keV&Y \\
      & 0005 & 36.6121 & -4.0741 & 3  & (4) &   &   &N  \\
         & 0006 & 36.6104 & -4.5286 & 3  & 12 & S &    &Y \\
\hline
0.30-0.40 (9973) & 0001 & 36.2974 & -4.7511 & 3  & (10) & A/M & &Y \\
         & 0002 & 36.6445 & -4.8804 & 4  & (11) & M &  &N\\
         & 0003 & 36.8090 & -4.6339 & 4  & (6) & G/M & &Y \\
         & 0004 & 36.6215 & -4.0689 & 4  & 4 &  XLSS014/W05/0.34 & &N  \\
\hline
0.35-0.45 (8248) & 0001 & 36.1413 & -4.8965 & 4  & 13 & A/M & &N \\
         & 0002 & 36.6542 & -4.9432 & 4  & (11) & G & &N \\
         & 0003 & 36.7937 & -4.6294 & 3  & (6) & G/M/S &&Y  \\
         & 0004 & 36.6156 & -4.0665 & 3  & (4) & S &  &N\\
\hline
0.40-0.50 (8302) & 0001 & 36.1201 & -4.8502 & 4  & (13) & A & &N\\
         & 0002 & 36.6735 & -4.9467 & 4  & (11) & G & &N \\
         & 0003 & 36.0884 & -4.0633 & 3  & (14) &  &  &N\\
         & 0004 & 36.6258 & -4.0680 & 3  & (4) & S & &N \\
\hline
0.45-0.55 (9216) & 0001 & 36.1152 & -4.8328 & 4  & (13) & A & &N\\
         & 0002 & 36.6799 & -4.0610 & 3  & (4) & G & &N\\
         & 0003 & 36.0798 & -4.0684 & 4  & 14 &  & &N \\
         & 0004 & 36.3842 & -4.2726 & 4  & 15 & S  &  &Y\\
         & 0005 & 36.8975 & -4.3768 & 3  & 16 &  &  &Y\\
         & 0006 & 36.8481 & -4.6202 & 3  & 17 & G/M & &Y \\
\hline
0.50-0.60 (10201) & 0001 & 36.1202 & -4.8557 & 4  & (13) & A &  &N\\
         & 0002 & 36.6990 & -4.0559 & 3  & (4) &  &  &N\\
         & 0003 & 36.0758 & -4.1999 & 4  & 18 & M &  &N\\
         & 0004 & 36.3762 & -4.2655 & 4  & (15) &  S  &  &Y\\
         & 0005 & 36.8334 & -4.4982 & 3  & (19) & M &  &Y\\
         & 0006 & 36.2840 & -4.7423 & 3  & 20 & A/M & &N\\
\hline
0.55-0.65 (10845) & 0001 & 36.1366 & -4.8951 & 3  & (13) & A & &N  \\
         & 0002 & 36.3825 & -4.2687 & 3  & (15) & S &  &Y\\
         & 0003 & 36.4645 & -4.4997 & 3  & 21 &  &  &Y\\
         & 0004 & 36.8646 & -4.5484 & 4  & 19 &  &  &Y\\
\hline
0.60-0.70 (10224) & 0001 & 36.6686 & -4.5096 & 4  & 22 & S &  &Y\\
         & 0002 & 36.2364 & -4.2232 & 3  & (23) &  &  &N\\
\hline
0.65-0.75 (9243) & 0001 & 36.0853 & -4.7942 & 3  & 24 & A & &N \\
         & 0002 & 36.4815 & -4.0820 & 3  & 25 & M/S & &N \\
         & 0003 & 36.7546 & -4.0752 & 3  & 26 &  &  &N\\
         & 0004 & 36.2089 & -4.2167 & 4  & 23 &  &  &N\\
         & 0005 & 36.6585 & -4.5160 & 4  & (22) & S &  &Y\\
\hline
\end{tabular}

\label{tab:candi}
\end{table*}

\begin{table*}
\caption[]{From central redshift of 0.75 to 1.05. We note that structure 35 
(general identification) is present from z=0.85 to z=1.25 with a small 
position shift on the sky. Structure 29 (general identification) is perhaps
identified with C3-c cluster (Pierre et al. 2006) but the case is not closed
as it is located very close to a masked region.}
\begin{tabular}{ccccccccc}
\hline
Slice &   id  & $\alpha$ & $\delta$ & Threshold in $\sigma$
& gen. id. & X-ray id and z XMM & T-X  & VVDS  \\
\hline
0.70-0.80 (9585) & 0001 & 36.9250 & -4.9052 & 3  & 27 & M & &N \\
         & 0002 & 36.1835 & -4.1749 & 4  & (23) &  &  &N\\
         & 0003 & 36.8791 & -4.2025 & 3  & 28 &  &  &Y\\
         & 0004 & 36.4924 & -4.4782 & 3  & (29) &  &  &Y\\
\hline
0.75-0.85 (9741) & 0001 & 36.0853 & -4.8695 & 3  & (30) & A &  &N\\
         & 0002 & 36.7614 & -4.0793 & 3  & 31 &  &  &N\\
         & 0003 & 36.1627 & -4.1807 & 4  & (23) &  &  &N\\
         & 0004 & 36.8789 & -4.2087 & 3  & (28) &  &  &Y\\
         & 0005 & 36.7091 & -4.2481 & 3  & 32 &  &  &N\\
         & 0006 & 36.5030 & -4.4714 & 3  & 29 & C3-c/0.82 ?  &  &Y\\
\hline
0.80-0.90 (11004) & 0001 & 36.0856 & -4.8648 & 4  & 30 & A &  &N\\
         & 0002 & 36.8458 & -4.1485 & 3  & (28) &  &  &Y\\
         & 0003 & 36.3781 & -4.1865 & 3  & 33 & S & &N\\
         & 0004 & 36.2113 & -4.2073 & 3  & (23) &  & &N \\
         & 0005 & 36.4003 & -4.4181 & 4  & (34) & &  &Y\\
\hline
0.85-0.95 (12019) & 0001 & 36.0864 & -4.8774 & 3  & (30) & A & &N \\
         & 0002 & 36.0587 & -4.2291 & 3  & (35) & M&  &N\\
         & 0003 & 36.3895 & -4.1940 & 4  & (33) & S &  &N\\
         & 0004 & 36.2141 & -4.2231 & 3  & (23) &  &  &N\\  
& 0005 & 36.3925 & -4.4135 & 4& 34&XLSSJ022534.2/A05/0.92  C3-b/0.92 & &Y \\
\hline
0.90-1.00 (11658) & 0001 & 36.0983 & -4.8933 & 4  & (30) & A &  &N\\
         & 0002 & 36.0664 & -4.2379 & 3  & (35) & M & &N \\
         & 0003 & 36.2245 & -4.2340 & 3  & (23) &  &  &N\\
         & 0004 & 36.2492 & -4.3086 & 4  & 36 &  & &N \\
         & 0005 & 36.3847 & -4.4106 & 3  & (34) & & &Y \\ 
\hline
0.95-1.05 (9998) & 0001 & 36.0989 & -4.9052 & 3  & (30) & A & &N\\
         & 0002 & 36.0872 & -4.2272 & 4  & 35 & E &  &N\\
         & 0003 & 36.2499 & -4.2590 & 4  & (23) &  &  &N\\
\hline
1.00-1.10 (8203) & 0001 & 36.7801 & -4.2833 & 3  & 37 & C1-005/1.05 & 3.7 keV &N\\
     & 0002 & 36.0875 & -4.1856 & 4  & (35) & E & &N\\
         & 0003 & 36.2513 & -4.2303 & 4  & (23) &  & &N\\
\hline
\end{tabular}

\label{tab:candi1}
\end{table*}

\begin{table*}
\caption{From a central redshift of 1.10 to 1.20.}
\begin{tabular}{ccccccccc}
\hline
Slice &  id  & $\alpha$ & $\delta$ & Threshold in $\sigma$
& gen. id. & X-ray id and z XMM & T-X   & VVDS   \\
\hline
1.05-1.15 (7755) & 0001 & 36.2635 & -4.2060 & 3  & (23) &  & &N \\
         & 0002 & 36.8640 & -4.2023 & 3  & 38 &  & &N \\
         & 0003 & 36.0848 & -4.2277 & 3  & (35) & M &  &N\\
\hline
1.10-1.20 (7612) & 0001 & 36.5663 & -4.8823 & 4  & 39 &  &  &N\\
         & 0002 & 36.9197 & -4.9233 & 3  & 40 & M & &N \\
         & 0003 & 36.0714 & -4.2046 & 4  & (35) & M &  &N\\
         & 0004 & 36.9054 & -4.1090 & 4  & 41 &  &  &N\\
\hline
1.15-1.25 (7662) & 0001 & 36.5183 & -4.8925 & 4  & (39) &  &  &N\\
         & 0002 & 36.9264 & -4.9257 & 4  & (40) & M & &N \\
         & 0003 & 36.0635 & -4.2135 & 4  & (35) & M &  &N\\
         & 0004 & 36.9042 & -4.4438 & 3  & 42 &  & &Y \\
         & 0005 & 36.8769 & -4.6289 & 4  & 43 & M/G & &Y \\
         & 0006 & 36.0565 & -4.0944 & 3  & 44 &  &  &N\\
\hline
\end{tabular}

\label{tab:candi2}
\end{table*}

\begin{table*}
\caption[]{Main structures detected and with VVDS spectroscopic data, id, 
coordinates, general identification, Nb 
of redshifts in the slice, Nb of redshifts in the system (when two systems are 
visible, we give both values), central redshift,  
velocity dispersion (when more than 4 available redshifts).  
We restrict here to VVDS redshift selected in the 
ellipse corresponding to every structure. When 
no value is given for the mean redshift and velocity dispersion, this means 
that the sparse sampling and/or the small number of data do not allow a 
significant characterization.}
\begin{tabular}{ccccccccc}
\hline
Slice &   id  & $\alpha$ & $\delta$ & gen. id. 
&N-slice& N-St & zcentral& $\sigma _v$ (km/s) \\
\hline
0.10-0.20 & 0001 & 36.3789 & -4.2424   & 1 & 3& 2 &0.138&\\
         & 0002 & 36.7981 & -4.1970   & 2 &           3  & 2    &0.185&\\
\hline
0.15-0.25 & 0001 & 36.3746 & -4.6831  & 3& 4 &2 &0.225&\\
         & 0003 & 36.6240 & -4.2523  & 5 &  11 & 5   &0.210& 258    \\
\hline
0.25-0.35& 0001 & 36.3166 & -4.7515 & 10 & 3& 3 &0.311&  \\
         & 0004 & 36.8416 & -4.5810 & 6 & 12& 6 &0.308& 391 \\
         & 0006 & 36.6104 & -4.5286 & 12 & 30  & 21       &0.313&727 \\
\hline
0.45-0.55& 0004 & 36.3842 & -4.2726  & 15& 2 &2  &0.542&\\
         & 0005 & 36.8975 & -4.3768 & 16& 2 &  1&0.53&\\
         & 0006 & 36.8481 & -4.6202 & 17& 4 &  3&0.543&\\
\hline
0.55-0.65& 0003 & 36.4645 & -4.4997 & 21 &15 &9  &0.613&594\\
         & 0004 & 36.8646 & -4.5484 & 19&10  &6  &0.610&864\\
\hline
0.60-0.70 & 0001 & 36.6686 & -4.5096 & 22&15  &7/6  &0.634/0.687&319/601\\
\hline
0.70-0.80& 0003 & 36.8791 & -4.2025 & 28 &2  & 2 &0.784&\\
\hline
0.75-0.85               & 0006 & 36.5030 & -4.4714 &29&   2  &  &&\\
\hline
0.85-0.95& 0005 & 36.3925 & -4.4135 & 34 &10&5 &0.920&488 \\
\hline
\end{tabular}
\label{tab:ZSPEC}
\end{table*}

Another remark concerns the detection of structures showing up
in several redshift slices. Two such structures (limiting ourselves to the non
heavily polluted by CFHTLS masking candidates and to z lower than 1.05:
structures 4 and 23) extend over redshift intervals 
of strictly more than 0.3 and are detected in each of the successive bins 
at the 3-$\sigma$ level (see Tab.~\ref{tab:fil}). This interval of 0.3 
represents $\pm$3 times the typical photometric redshift uncertainty. It is
also also larger than the catastrophic errors. This ensures us that we are 
probably not dealing with artefacts. It can still be chance alignements of 
real structures. However, if not, this is really a puzzling fact as the 
length of these filaments (or structure chains) is several hundreds of Mpc! 
Their radial extension is clearly larger than the maximal void sizes computed
in Hoyle $\&$ Vogeley (2004). If these filaments are real, then they have to
cross at least one node (the place where the massive clusters form) of the
cosmic web and to percolate from a cosmic cell to another one. We should 
therefore detect massive clusters inside these filaments in the XMM-LSS data
and these are, indeed, associated with X-ray structures. 

\begin{table}
\caption[]{Redshift detection interval of the radial filaments, mean 
coordinates, X-ray association, redshift 
extension in Mpc and redshift extension in Mpc quadratically diminished 
with the $\pm$1$\sigma$ photometric redshift uncertainty at the given 
redshift.}
\begin{tabular}{cccccc}
\hline
z det. int. & $\alpha$ & $\delta$ & X-ray & red. ext. & corr. ext.\\
\hline
1.15 - 0.60 & 36.25 & -4.20 & yes & 1467 Mpc & 1421 Mpc \\
0.60 - 0.15 & 36.60 & -4.05 & yes & 1632 Mpc & 1581 Mpc \\
\hline
\end{tabular}
\label{tab:fil}
\end{table}

\section{Conclusions}

We show in this paper that using the excellent quality photometric redshifts
computed on the D1 CFHTLS field by Ilbert et al. (2006) and combining them
with an adaptative kernel galaxy density estimate, we are able to efficiently
detect structures up to z$\sim$1.05 without any hypotheses on the nature of 
what the structures are. 

The analysis based on 
slices in redshift space allows to reduce efficiently fore and background 
contamination, then increasing the contrast of real structures. 

Our detections, taking into account biases of both analyses, are in good
agreement with X-ray detections (and sometimes help to recover them) and also
allow to detect low mass structures, invisible for X-ray surveys. Detections 
with no evident X-ray counterpart are in general confirmed by spectroscopic
information when available. The efficiency of the method seems to be due also  to the
fact that  light  appears to trace mass in clusters  which has been verified
at least for small redshifts (e.g 
Katgert et al. 2004). It is then encouraging in this perspective to use our
method in parallel with others to count clusters both in simulations with
realistic galaxy representation and in the real universe.   
We detect at least two structure-chains of several hundreds of Mpc (structures
4 and 23). The size of the D1 field is, however, far too small to 
conduce quantitative cosmological studies, but it allows to calibrate our 
method. Such quantitative studies will be achieved in future works using 
other large scale and deep CFHTLS fields.

\begin{acknowledgements}
The authors thanks C. Benoist and L.F. Olsen for useful discussions.
This work is based in part on data obtained with 
the European Southern Observatory on Paranal, Chile, and on
observations obtained with MegaPrime/Megacam, a joint project
of CFHT and CEA/DAPNIA, at the Canada France Hawaii Telescope (CFHT)
which is operated by the National Research Council (NRC) of Canada,
the Institut National des Science de l'Univers of the Centre
National de la Recherche Scientifique (CNRS) of France, and the
University of Hawaii. This work is based in part on data products
produced at TERAPIX and the Canadian Astronomy Data Centre as
part of the Canada-France-Hawaii Telescope Legacy Survey, a
collaborative project of NRC and CNRS.
This work is based in part on observations obtained with XMM-Newton, an ESA
science mission with instruments and contributions directly funded by ESA
Member states and NASA.
The VLT-VIMOS observations have been carried out on guaranteed time (GTO) 
allocated by the European Southern Observatory (ESO) to the VIRMOS consortium, 
under a contractual agreement between the Centre National de la Recherche 
Scientifique of France, heading a consortium of French and Italian 
institutes, and ESO, to design, manufacture and test the VIMOS instrument.
\end{acknowledgements}

\Online

\begin{appendix} 

\begin{table*}
\caption{Same as Tabs.~\ref{tab:candi} to ~\ref{tab:candi2} for slices with 
central redshifts between 1.25 to 1.425.}
\begin{tabular}{cccccccccc}
\hline
Slice &   Structure id  & $\alpha$ & $\delta$ & Threshold in $\sigma$
& gen. id. & X-ray id and z XMM & T-X   & VVDS & Nb  \\
\hline
1.20-1.30 (7174) & 0001 & 36.9239 & -4.9189 & 4  & (40) & M & &N& \\
         & 0002 & 36.4878 & -4.9125 & 4  & (39) &  & &N& \\
         & 0003 & 36.9052 & -4.4377 & 4  & (42) &  &  &Y&1\\
         & 0004 & 36.8669 & -4.6201 & 4  & (43) & M/G &  &N&\\
         & 0005 & 36.0596 & -4.0896 & 3  & (44) &  &  &N&\\
\hline
1.25-1.35 (6352) & 0001 & 36.7340 & -4.7891 & 4  & 45 &   &  &N&\\
         & 0002 & 36.0588 & -4.1025 & 3  & (44) &  &  &N&\\
         & 0003 & 36.8989 & -4.4477 & 4  & (42) &  &  &N&\\
         & 0004 & 36.8744 & -4.6244 & 3  & (43) & M/G &  &N&\\
\hline
1.30-1.40 (5004) & 0001 & 36.7386 & -4.7950 & 4  & (45) &  & &N& \\
         & 0002 & 36.8913 & -4.4610 & 3  & (42) &  &  &Y&1\\
\hline
1.35-1.50 (5673) & 0001 & 36.8357 & -4.6087 & 3  & 46 & M/G &  &N&\\
         & 0002 & 36.3528 & -4.1316 & 3  & 47 & S & &N& \\
         & 0003 & 36.9115 & -4.4723 & 3  & (42) &  & &N& \\
\hline
\end{tabular}
\label{tab:candiup}
\end{table*}

\end{appendix}

\end{document}